\documentclass[aps,prl,nofootinbib,superscriptaddress,reprint]{revtex4-1}
\usepackage{graphicx}
\usepackage{amsmath}
\usepackage{units}
\usepackage{amssymb}

\begin{document}

\title{Search for photon line-like signatures from Dark Matter annihilations with H.E.S.S.}
\date{\today}

\author{A.~Abramowski}
\affiliation{Universit\"at Hamburg, Institut f\"ur Experimentalphysik, Luruper Chaussee 149, D 22761 Hamburg, Germany}
\author{F.~Acero}
\affiliation{Laboratoire Univers et Particules de Montpellier, Universit\'e Montpellier 2, CNRS/IN2P3,  CC 72, Place Eug\`ene Bataillon, F-34095 Montpellier Cedex 5, France}
\author{F.~Aharonian}
\affiliation{Max-Planck-Institut f\"ur Kernphysik, P.O. Box 103980, D 69029 Heidelberg, Germany}
\affiliation{Dublin Institute for Advanced Studies, 31 Fitzwilliam Place, Dublin 2, Ireland}
\affiliation{National Academy of Sciences of the Republic of Armenia, Yerevan }
\author{A.G.~Akhperjanian}
\affiliation{Yerevan Physics Institute, 2 Alikhanian Brothers St., 375036 Yerevan, Armenia}
\affiliation{National Academy of Sciences of the Republic of Armenia, Yerevan }
\author{G.~Anton}
\affiliation{Universit\"at Erlangen-N\"urnberg, Physikalisches Institut, Erwin-Rommel-Str. 1, D 91058 Erlangen, Germany}
\author{S.~Balenderan}
\affiliation{University of Durham, Department of Physics, South Road, Durham DH1 3LE, U.K.}
\author{A.~Balzer}
\affiliation{DESY, D-15735 Zeuthen, Germany}
\affiliation{Institut f\"ur Physik und Astronomie, Universit\"at Potsdam,  Karl-Liebknecht-Strasse 24/25, D 14476 Potsdam, Germany}
\author{A.~Barnacka}
\affiliation{Nicolaus Copernicus Astronomical Center, ul. Bartycka 18, 00-716 Warsaw, Poland}
\affiliation{CEA Saclay, DSM/Irfu, F-91191 Gif-Sur-Yvette Cedex, France}
\author{Y.~Becherini}
\affiliation{APC, AstroParticule et Cosmologie, Universit\'{e} Paris Diderot, CNRS/IN2P3, CEA/Irfu, Observatoire de Paris, Sorbonne Paris Cit\'{e}, 10, rue Alice Domon et L\'{e}onie Duquet, 75205 Paris Cedex 13, France, }
\affiliation{Laboratoire Leprince-Ringuet, Ecole Polytechnique, CNRS/IN2P3, F-91128 Palaiseau, France}
\author{J.~Becker~Tjus}
\affiliation{Institut f\"ur Theoretische Physik, Lehrstuhl IV: Weltraum und Astrophysik, Ruhr-Universit\"at Bochum, D 44780 Bochum, Germany}
% ...

\collaboration{H.E.S.S. Collaboration}\noaffiliation

\begin{abstract}

Gamma-ray line signatures can be expected in the very-high-energy (VHE; $ E_\gamma > 100$~GeV) domain due to self-annihilation or decay of dark matter (DM) particles in space. Such a signal would be readily distinguishable from astrophysical $\gamma$-ray sources that in most cases produce continuous spectra which span over several orders of magnitude in energy. Using data collected with the H.E.S.S. $\gamma$-ray instrument, upper limits on line-like emission are obtained in the energy range between $\sim\unit[500]{GeV}$ and $\sim\unit[25]{TeV}$ for the central part of the Milky Way halo and for extragalactic observations, complementing recent limits obtained with the Fermi-LAT instrument at lower energies. No statistically significant signal could be found. For monochromatic $\gamma$-ray line emission, flux limits of $\unit[(2\times10^{-7}$ -- $2\times10^{-5})]{m^{-2} s^{-1} sr^{-1}}$ and $\unit[(1\times10^{-8}$ -- $2\times10^{-6})]{m^{-2} s^{-1} sr^{-1}}$ are obtained for the central part of the Milky Way halo and extragalactic observations, respectively.
For a DM particle mass of \unit[1]{TeV}, limits on the velocity-averaged DM annihilation cross section $\langle \sigma v\rangle_{\chi\chi\to\gamma\gamma}$ reach $\sim\unit[10^{-27}]{cm^{3} s^{-1}}$, based on the Einasto parametrization of the Galactic DM halo density profile.

\end{abstract}

\maketitle
\section{Introduction}

In the last few years, imaging atmospheric Cherenkov telescopes (IACTs) have been used to search for dark matter (DM) signals in very-high-energy (VHE; $E_\gamma > 100$~GeV) $\gamma$~rays \cite{Aharonian:2008dm,Wood:2008,Aharonian:2010dmerr,Albert:2008,Aharonian:2009dm,Veritas:2010dm,%
Aharonian:2006wh,NekrassovEldik:2010,MagicSegue:2011,VeritasSegue:2012}. Objects with large predicted DM density, like the Galactic centre (GC), the central Galactic halo region (CGH), dwarf galaxies or centres of nearby galaxies were studied. All such searches concentrated on the detection of $\gamma$~rays produced in decays of secondary particles -- mostly neutral mesons -- in the process of DM self-annihilation or decay (see, e.\,g., \cite{Hill:1987, Tasitsiomi:2002}). The broad energy distribution of such $\gamma$~rays is continuous and therefore more difficult to distinguish from $\gamma$-ray emission from astrophysical (particle accelerating) sources, as opposed to spectral features, which would pose a much more striking evidence for a DM-induced $\gamma$-ray signal. The most prominent spectral feature is a $\gamma$-ray line\footnote{Note, however, that VHE $\gamma$-ray line features may also arise due to unshocked $e^+/e^-$-winds created by pulsars \cite{PulsarLines:2000}.}, which, for DM self-annihilation into $\gamma \gamma / \gamma Z$ (and $m_\chi \gg m_\mathrm{Z}$), is expected at an energy at or close to the DM particle mass, $E_\gamma \approx m_{\chi}$. For a decay $\chi\rightarrow \gamma X$ of a DM particle $\chi$ with $m_\chi \gg m_\mathrm{X}$, $E_\gamma \approx m_{\chi}/2$.
Such annihilations or decays are, however, loop-suppressed, since electrically neutral DM particles do not couple to photons directly. Nonetheless, recent theoretical developments show the possibility of a rather pronounced spectral feature for some implementations of particle physics beyond the Standard Model (see, e.\,g., \cite{Bergstroem:2007}). In these models, spectral signatures may arise due to the radiation of a hard photon from real or virtual charged particles created in the annihilation process and be a dominant component in the overall $\gamma$-ray annihilation spectrum. Here a search for $\gamma$-ray line-like signatures conducted with the H.E.S.S. experiment in the energy range $E_\gamma \sim \unit[500]{GeV} - \unit[25]{TeV}$ is reported, complementing a recent search at energies between \unit[7]{GeV} and \unit[200]{GeV} with the Fermi-LAT instrument \cite{FermiLines:2012} and studies discussing an indication for a line feature at an energy of about \unit[130]{GeV} \cite{Bringmann:2012,Weniger:2012,Tempel:2012}.

The search for a DM-induced spectral signature in the H.E.S.S. data is performed separately for two sky regions of interest. The first is the CGH, a promising region due to its proximity and predicted large DM concentration. Following \cite{NekrassovEldik:2010}, the search region is defined as a circle of $1^\circ$ radius centred on the GC, where the Galactic plane is excluded, by requiring $|b| > 0.3^\circ$. The second region is the extragalactic sky covered by H.E.S.S. observations, with regions containing known VHE $\gamma$-ray sources being excluded from the analysis. 
For both data sets, the uncertainty on the strength of a putative DM annihilation signal is much reduced in comparison to the observations of centres of galaxies:
for the CGH, the very centre is not considered, thus avoiding a region where the DM profile is only poorly constrained \cite{NekrassovEldik:2010}. For the extragalactic data set, differences in DM density between individual substructures are averaged out by observing many different fields of view \cite{Bergstrom:2001}. One should note, however, that a potentially large (but highly uncertain) $\gamma$-ray flux from Galactic DM annihilations may contribute to the extragalactic analysis \cite{Pieri:2011}.

\section{Methodology and results}

The CGH data set is composed of $\unit[112]{h}$ (live time) of GC observations recorded with the H.E.S.S. VHE $\gamma$-ray instrument (see \cite{Crab:2006} and references therein) during the years 2004--2008\footnote{Data from later periods were excluded, since the gradual degradation in time of the optical efficiency of the instrument would result in an increased energy threshold.}. The mean distance between the telescope pointing positions and the GC is $0.7^\circ$, with a maximum of $1.5^\circ$ \cite{NekrassovEldik:2010}. The extragalactic data set comprises $\unit[1153]{h}$ of H.E.S.S. observations taken during 2004--2007, targeted at various extragalactic objects. Regions in the field-of-view (FoV) containing known VHE $\gamma$-ray sources are excluded by masking out a circular region (of radius 0.2$^\circ$ for point sources) around the source position. 

Observations with zenith angles larger than $30^\circ$ are excluded from the analysis to lower the energy threshold, resulting in a mean zenith angle of $14^\circ$ ($19^\circ$) for the CGH (extragalactic) observations. Only $\gamma$-ray-like events are accepted for which the distance between the reconstructed $\gamma$-ray direction and the observation direction of the H.E.S.S. array is smaller than $2^\circ$, avoiding showers being reconstructed too close to the edges of the $\sim 5^\circ$ diameter FoV of the H.E.S.S. cameras \cite{Crab:2006}. Furthermore, events are considered only if they pass H.E.S.S. standard $\gamma$-ray selection criteria defined in \cite{Crab:2006} and triggered all four telescopes. Only $\unit[15]{\%}$ of the total event sample is kept by the latter selection. However, compared to the H.E.S.S. standard analysis, such selection leads to a better signal to background ratio and an improved energy resolution of Gaussian width $\sigma_\mathrm{E}$ ($\unit[17]{\%}$ at $\unit[500]{GeV}$ and $\unit[11]{\%}$ at $\unit[10]{TeV}$), and therefore increases the sensitivity of the analysis to spectral features by up to $50\%$. The energy threshold is $\unit[310]{GeV}$ ($\unit[500]{GeV}$) for the CGH (the extragalactic) data set.

\begin{figure}
\includegraphics[width=0.5\textwidth]{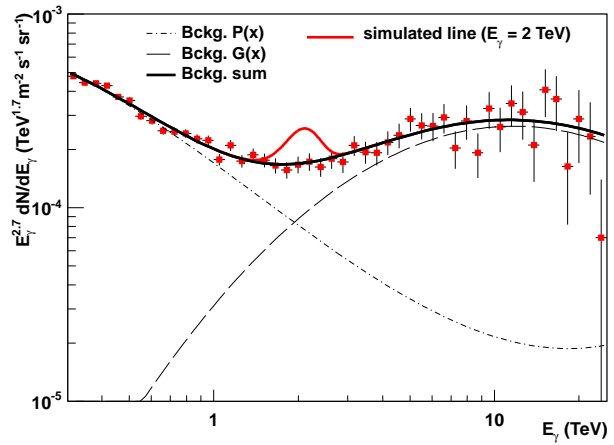}
\caption{Reconstructed flux spectrum of the CGH region, using 25 equidistant bins per unit of $\log_{10}(E_\gamma)$. Flux points have been multiplied by $E_\gamma^{2.7}$. The data consist mostly of hadronic cosmic ray background events, reconstructed using a $\gamma$-ray hypothesis. The spectrum is well described by the parametrization introduced in Eq.~\ref{Eq:BGFunc}, depicted by the black solid line. The corresponding $\chi^2$-test probability is $p=0.34$. The two contributions $P(x)$ and $G(x)$ are shown by the dashed-dotted and the dashed curve, respectively. Note that the shape of the Gaussian function $G(x)$ is much broader than the expected monochromatic line feature from DM annihilations. As an example, the red curve shows the expected signal of a line at $E_\gamma=\unit[2]{TeV}$ that would be detected with a statistical significance of 5 standard deviations above the background.}
\label{Fig:SpectrumWithFunc}
\end{figure}

Differential flux spectra are calculated from the reconstructed event energies separately for the CGH and extragalactic data sets  using zenith angle-, energy- and offset-dependent effective collection areas from $\gamma$-ray simulations.
Since sky regions containing known VHE $\gamma$-ray sources were excluded from the analysis, the spectra consist mostly of $\gamma$-ray-like cosmic-ray background events (and a fraction of $\sim 10\%$ of electrons). These spectra are well described by the empirical parametrization
\begin{equation}
\frac{\mathrm{d}N}{\mathrm{d}E_\gamma} = a_0 \left(\frac{E_\gamma}{\unit[1]{TeV}}\right)^{-2.7} \left[P(x) + \beta G(x)\right],
\label{Eq:BGFunc}
\end{equation}
where $E_\gamma$ is the reconstructed energy of the event under $\gamma$-ray hypothesis and $P(x) = \exp{(a_1 x + a_2 x^2 + a_3 x^3)}$. $G(x)$ is a Gaussian function with mean $\mu_\mathrm{x}$ and rms $\sigma_\mathrm{x}$, and $x = \log_{10}(E_\gamma/\unit[1]{TeV})$.  The free parameters $a_{0\dots 3}$, $\beta$, $\mu_\mathrm{x}$, and $\sigma_\mathrm{x}$ are optimized simultaneously by a maximum likelihood approach based on the binned event count spectrum. Since the number of reconstructed counts $n_i$ in energy bin $i$ of the count spectrum is Poisson-distributed, the log-likelihood function takes the form
\begin{equation*}
\ln\mathcal{L} = \sum_{i=1}^N n_i\ln\lambda_i - \lambda_i,
\end{equation*}
where $\lambda_i$ is the number of counts in bin $i$ that is expected according to the flux spectrum parametrization given in Eq.~\ref{Eq:BGFunc}, and $N$ is the total number of bins of the count spectrum. As an example, Fig.~\ref{Fig:SpectrumWithFunc} shows the differential flux spectrum and the best-fit background parametrization obtained for the CGH data set.

On top of the smooth cosmic ray flux spectrum, a monochromatic $\gamma$-ray line\footnote{In this context, the term 'monochromatic line' refers to spectral features with energy width much smaller than the energy resolution $\sigma_\mathrm{E}$ of the H.E.S.S.\ instrument.} may be identified as a Gaussian peak of width $\sigma_\mathrm{E}$ centred at the line energy $E_\gamma$. To search for such lines, a Gaussian term with fixed energy $E_\gamma$ and fixed corresponding width $\sigma_\mathrm{E}$ was added to the spectrum parametrization given in Eq.~\ref{Eq:BGFunc}. The spectrum was refit, and from the normalization of the Gaussian the flux of the putative line was reconstructed. By repeating this procedure, using ten logarithmically equidistant energies $E_\gamma$ per decade of energy, the flux spectrum was scanned for monochromatic $\gamma$-ray signatures. Line scans were performed in the energy range $\unit[0.5]{TeV}$--$\unit[20]{TeV}$ and $\unit[0.8]{TeV}$--$\unit[25]{TeV}$ for the CGH and the extragalactic data sets, respectively. 

\begin{figure}
\includegraphics[width=0.5\textwidth]{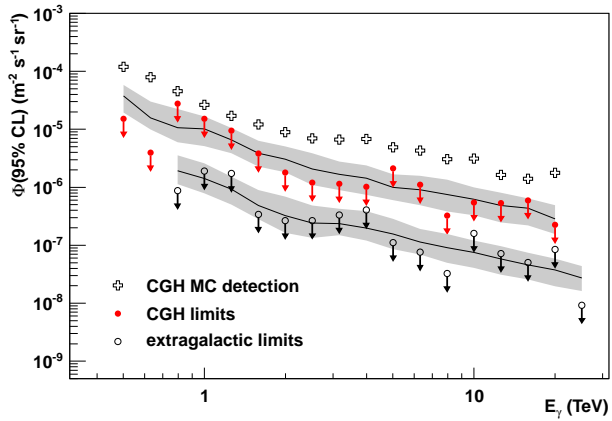}
\caption{Upper limits on $\gamma$-ray flux from monochromatic line signatures, derived from the CGH region (red arrows with full data points) and from extragalactic observations (black arrows with open data points). For both data sets, the solid black lines show the mean expected limits derived from a large number of statistically randomized simulations of fake background spectra, and the gray bands denote the corresponding 68\% CL regions for these limits. Black crosses denote the flux levels needed for a statistically significant line detection in the CGH dataset.}
\label{Fig:FluxLimits}
\end{figure}

No $\gamma$-ray line flux was found to exceed the \textit{a-priori} chosen detection threshold of $\Delta\ln\mathcal{L} = 12.5$, corresponding to a significance of 5 standard deviations above the background level for Gaussian parameters. Thus flux upper limits were calculated by constraining the flux normalization of the Gaussian to be non-negative in the fit and using the MINOS package from the Minuit\cite{James:1975dr} fitting tool to calculate asymmetric errors with error level $\Delta\ln\mathcal{L} = 1.35$, corresponding to a 95\%~CL one-sided limit on the flux of the line \cite{James:2004, FermiLines:2012}. These limits are shown in Fig.~\ref{Fig:FluxLimits}. To test whether the limits are compatible with random fluctuations of the background, a large number of statistically randomized fake background spectra was simulated using the best-fit background parametrization as an input, and limits were obtained for each of these spectra. The resulting mean limits, together with the 68\% CL region calculated from the limit distribution at each test energy, are shown in Fig.~\ref{Fig:FluxLimits} for comparison. Also shown are mean reconstructed fluxes from simulated lines that are detected with a significance of 5 standard deviations using the above prescription.

\begin{figure}
\includegraphics[width=0.5\textwidth]{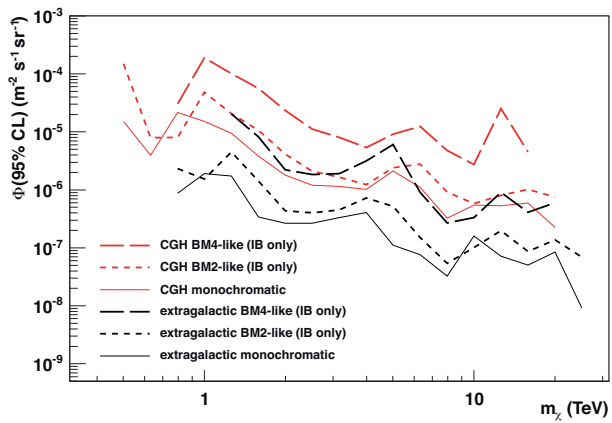}
\caption{Flux upper limits on spectral features arising from the emission of a hard photon in the DM annihilation process. Limits are exemplary shown for features of comparable shape to those arising in the models BM2 and BM4 given in \cite{Bergstroem:2007}.
The monochromatic line limits, assuming $m_\chi=E_\gamma$, are shown for comparison.}
\label{Fig:BMLimits}
\end{figure}

Additionally, flux upper limits were determined for broader spectral features like those arising due to internal bremsstrahlung (IB). As an example, calculations by \cite{Bergstroem:2007} in the framework of supersymmetric models predict the contribution of IB photons to the $\gamma$-ray spectrum to dominate over secondary $\gamma$-ray production for photon energies close to the DM (neutralino) mass $m_\chi$. 
Flux upper limits for the benchmark models BM2 and BM4 of \cite{Bergstroem:2007} were calculated following the technique described above. Firstly, the signal shapes predicted by the models were convolved with the energy response of the instrument. Together with the background parametrization, the resulting templates were then fitted (with the normalization of the template and the background parameters being free variables in the fit) to the flux spectrum. Note that only the IB part of the full annihilation spectra of these models is considered since the contribution from production of secondary photons steeply decreases towards $m_\chi$ (see \cite{Bergstroem:2007}), and is therefore hard to discriminate against the cosmic-ray background. In any case, since these models were calculated for a very specific set of MSSM parameters (and hence neutralino mass), they can only serve as a template to demonstrate the sensitivity of H.E.S.S.\ for features of similar shape (and are therefore referred to as BM2-like and BM4-like limits). Fig.~\ref{Fig:BMLimits} shows that -- because of the intrinsic widths of the expected features -- these limits are typically weaker by a factor two (BM2-like) to ten (BM4-like) compared to the monochromatic line limits. Note that all flux limits do also constrain putative features in the spectrum of cosmic ray electrons and positrons, since the H.E.S.S. experiment exhibits a similar sensitivity for detecting these particles as for $\gamma$~rays.

Possible systematic uncertainties due to the unknown shape of the background spectrum have been extensively studied, e.\,g.\ by changing the background parametrization described in Eq.~\ref{Eq:BGFunc} to one based on Legendre polynomials. The background parametrization does not show any significant correlation with shape parameters of spectral signatures, in particular with regard to the $G(x)$ term. The stability of the $\gamma$-ray flux reconstruction was investigated by adding artificial peaks to the background spectrum and reconstructing them with the fitting procedure described above. The systematic uncertainty on the reconstructed peak flux was of the order of a few percent, and the fit of the background was found to be very stable and independent of the location and normalization of the artificial peak. On the other hand, despite detailed Monte-Carlo simulations of the instrument, the true energy resolution $\sigma_\mathrm{E}$ of the instrument might be underestimated. When $\sigma_\mathrm{E}$ is artificially enlarged by e.\,g.\ \unit[20]{\%} -- i.\,e.\ $\sigma_\mathrm{E} = \unit[20]{\%}\  (\unit[13]{\%})$ at $E_\gamma=\unit[500]{GeV}\ (\unit[10]{TeV})$ --, upper limits get shifted to larger values by about \unit[15--20]{\%}, depending on the energy and the statistics in the individual spectrum bins. The maximum shift is observed in the extragalactic limit curve and amounts to \unit[40]{\%}. In total, the systematic error on the flux upper limits is estimated to be about $\unit[50]{\%}$. All flux upper limits were cross-checked using an alternative analysis framework \cite{Naurois:2009}, with an independent calibration of camera pixel amplitudes, and a different event reconstruction and event selection method, leading to results well consistent within the quoted systematic error. 

\begin{figure}
\includegraphics[width=0.5\textwidth]{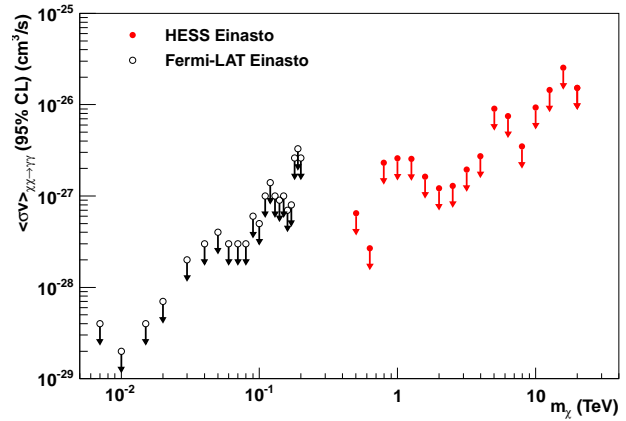}
\caption{Limits on the velocity-weighted cross section for DM annihilation into two photons calculated from the CGH flux limits (red arrows with full data points). The Einasto density profile with parameters described in \cite{Pieri:2011} was used. Limits obtained by Fermi-LAT, assuming the Einasto profile as well, are shown for comparison (black arrows with open data points) \cite{FermiLines:2012}.}
\label{Fig:CSLimits}
\end{figure}

For the Einasto parametrization of the DM density distribution in the Galactic halo \cite{Pieri:2011}, limits on the velocity-weighted DM annihilation cross section into $\gamma$ rays, $\langle \sigma v \rangle_{\chi\chi\to\gamma\gamma}$, are calculated from the CGH flux limits using the astrophysical factors given in \cite{NekrassovEldik:2010}. The result is shown in Fig.~\ref{Fig:CSLimits} and compared to recent results obtained at GeV energies with the Fermi-LAT instrument.

\section{Summary and Conclusions}

For the first time, a search for spectral $\gamma$-ray signatures at very-high energies was performed based on H.E.S.S. observations of the central Milky Way halo region and extragalactic sky. Both regions of interest exhibit a reduced dependency of the putative DM annihilation flux on the actual DM density profile. 
Upper limits on monochromatic $\gamma$-ray line signatures were determined for the first time for energies between $\sim\unit[500]{GeV}$ and $\sim\unit[25]{TeV}$, covering an important region of the mass range of particle DM. Additionally, limits were obtained on spectral signatures arising from internal bremsstrahlung processes, as 
predicted by the models BM2 and BM4 of \cite{Bergstroem:2007}. It should be stressed that the latter results are valid for all spectral signatures of comparable shape. Besides, all limits also apply for potential signatures in the spectrum of cosmic-ray electrons and positrons. 

Flux limits on monochromatic line emission from the central Milky Way halo were used to calculate upper limits on $\langle \sigma v \rangle_{\chi\chi\to\gamma\gamma}$. Limits are obtained in a neutralino mass range that is complementary to the line searches performed by Fermi-LAT \cite{FermiLines:2012}, reaching $\sim\unit[10^{-27}]{cm^{3}  s^{-1}}$ at a DM mass of \unit[1]{TeV}, comparable to those obtained by Fermi-LAT at energies of  $\sim\unit[100]{GeV}$.

\begin{acknowledgements}
The support of the Namibian authorities and of the University of Namibia
in facilitating the construction and operation of H.E.S.S. is gratefully
acknowledged, as is the support by the German Ministry for Education and
Research (BMBF), the Max Planck Society, the French Ministry for Research,
the CNRS-IN2P3 and the Astroparticle Interdisciplinary Programme of the
CNRS, the U.K. Science and Technology Facilities Council (STFC),
the IPNP of the Charles University, the Polish Ministry of Science and 
Higher Education, the South African Department of
Science and Technology and National Research Foundation, and by the
University of Namibia. We appreciate the excellent work of the technical
support staff in Berlin, Durham, Hamburg, Heidelberg, Palaiseau, Paris,
Saclay, and in Namibia in the construction and operation of the
equipment.
\end{acknowledgements}

\bibliography{LinesPaper}

\begin{thebibliography}{24}%
\makeatletter
\providecommand \@ifxundefined [1]{%
 \@ifx{#1\undefined}
}%
\providecommand \@ifnum [1]{%
 \ifnum #1\expandafter \@firstoftwo
 \else \expandafter \@secondoftwo
 \fi
}%
\providecommand \@ifx [1]{%
 \ifx #1\expandafter \@firstoftwo
 \else \expandafter \@secondoftwo
 \fi
}%
\providecommand \natexlab [1]{#1}%
\providecommand \enquote  [1]{``#1''}%
\providecommand \bibnamefont  [1]{#1}%
\providecommand \bibfnamefont [1]{#1}%
\providecommand \citenamefont [1]{#1}%
\providecommand \href@noop [0]{\@secondoftwo}%
\providecommand \href [0]{\begingroup \@sanitize@url \@href}%
\providecommand \@href[1]{\@@startlink{#1}\@@href}%
\providecommand \@@href[1]{\endgroup#1\@@endlink}%
\providecommand \@sanitize@url [0]{\catcode `\\12\catcode `\$12\catcode
  `\&12\catcode `\#12\catcode `\^12\catcode `\_12\catcode `\%12\relax}%
\providecommand \@@startlink[1]{}%
\providecommand \@@endlink[0]{}%
\providecommand \url  [0]{\begingroup\@sanitize@url \@url }%
\providecommand \@url [1]{\endgroup\@href {#1}{\urlprefix }}%
\providecommand \urlprefix  [0]{URL }%
\providecommand \Eprint [0]{\href }%
\providecommand \doibase [0]{http://dx.doi.org/}%
\providecommand \selectlanguage [0]{\@gobble}%
\providecommand \bibinfo  [0]{\@secondoftwo}%
\providecommand \bibfield  [0]{\@secondoftwo}%
\providecommand \translation [1]{[#1]}%
\providecommand \BibitemOpen [0]{}%
\providecommand \bibitemStop [0]{}%
\providecommand \bibitemNoStop [0]{.\EOS\space}%
\providecommand \EOS [0]{\spacefactor3000\relax}%
\providecommand \BibitemShut  [1]{\csname bibitem#1\endcsname}%
\let\auto@bib@innerbib\@empty
\bibitem [{\citenamefont {{Aharonian}}\ \emph {et~al.}(2008)\citenamefont
  {{Aharonian}} \emph {et~al.}}]{Aharonian:2008dm}%
  \BibitemOpen
  \bibfield  {author} {\bibinfo {author} {\bibfnamefont {F.}~\bibnamefont
  {{Aharonian}}} \emph {et~al.} (\bibinfo {collaboration} {H.E.S.S.
  Collaboration}),\ }\href {\doibase 10.1016/j.astropartphys.2007.11.007}
  {\bibfield  {journal} {\bibinfo  {journal} {Astropart. Phys.}\ }\textbf
  {\bibinfo {volume} {29}},\ \bibinfo {pages} {55} (\bibinfo {year}
  {2008})}\BibitemShut {NoStop}%
\bibitem [{\citenamefont {{Wood}}\ \emph {et~al.}(2008)\citenamefont {{Wood}}
  \emph {et~al.}}]{Wood:2008}%
  \BibitemOpen
  \bibfield  {author} {\bibinfo {author} {\bibfnamefont {M.}~\bibnamefont
  {{Wood}}} \emph {et~al.},\ }\href {\doibase 10.1086/529421} {\bibfield
  {journal} {\bibinfo  {journal} {\apj}\ }\textbf {\bibinfo {volume} {678}},\
  \bibinfo {pages} {594} (\bibinfo {year} {2008})}\BibitemShut {NoStop}%
\bibitem [{\citenamefont {{Aharonian}}\ \emph {et~al.}(2010)\citenamefont
  {{Aharonian}} \emph {et~al.}}]{Aharonian:2010dmerr}%
  \BibitemOpen
  \bibfield  {author} {\bibinfo {author} {\bibfnamefont {F.}~\bibnamefont
  {{Aharonian}}} \emph {et~al.} (\bibinfo {collaboration} {H.E.S.S.
  Collaboration}),\ }\href {\doibase 10.1016/j.astropartphys.2010.01.007}
  {\bibfield  {journal} {\bibinfo  {journal} {Astropart. Phys.}\ }\textbf
  {\bibinfo {volume} {33}},\ \bibinfo {pages} {274} (\bibinfo {year}
  {2010})}\BibitemShut {NoStop}%
\bibitem [{\citenamefont {{Albert}}\ \emph {et~al.}(2008)\citenamefont
  {{Albert}} \emph {et~al.}}]{Albert:2008}%
  \BibitemOpen
  \bibfield  {author} {\bibinfo {author} {\bibfnamefont {J.}~\bibnamefont
  {{Albert}}} \emph {et~al.},\ }\href {\doibase 10.1086/529135} {\bibfield
  {journal} {\bibinfo  {journal} {\apj}\ }\textbf {\bibinfo {volume} {679}},\
  \bibinfo {pages} {428} (\bibinfo {year} {2008})}\BibitemShut {NoStop}%
\bibitem [{\citenamefont {{Aharonian}}\ \emph {et~al.}(2009)\citenamefont
  {{Aharonian}} \emph {et~al.}}]{Aharonian:2009dm}%
  \BibitemOpen
  \bibfield  {author} {\bibinfo {author} {\bibfnamefont {F.}~\bibnamefont
  {{Aharonian}}} \emph {et~al.} (\bibinfo {collaboration} {H.E.S.S.
  Collaboration}),\ }\href {\doibase 10.1088/0004-637X/691/1/175} {\bibfield
  {journal} {\bibinfo  {journal} {\apj}\ }\textbf {\bibinfo {volume} {691}},\
  \bibinfo {pages} {175} (\bibinfo {year} {2009})}\BibitemShut {NoStop}%
\bibitem [{\citenamefont {{Acciari}}\ \emph {et~al.}(2010)\citenamefont
  {{Acciari}} \emph {et~al.}}]{Veritas:2010dm}%
  \BibitemOpen
  \bibfield  {author} {\bibinfo {author} {\bibfnamefont {V.~A.}\ \bibnamefont
  {{Acciari}}} \emph {et~al.},\ }\href {\doibase 10.1088/0004-637X/720/2/1174}
  {\bibfield  {journal} {\bibinfo  {journal} {\apj}\ }\textbf {\bibinfo
  {volume} {720}},\ \bibinfo {pages} {1174} (\bibinfo {year}
  {2010})}\BibitemShut {NoStop}%
\bibitem [{\citenamefont {Aharonian}\ \emph
  {et~al.}(2006{\natexlab{a}})\citenamefont {Aharonian} \emph
  {et~al.}}]{Aharonian:2006wh}%
  \BibitemOpen
  \bibfield  {author} {\bibinfo {author} {\bibfnamefont {F.}~\bibnamefont
  {Aharonian}} \emph {et~al.} (\bibinfo {collaboration} {H.E.S.S.
  Collaboration}),\ }\href@noop {} {\bibfield  {journal} {\bibinfo  {journal}
  {Phys. Rev. Lett.}\ }\textbf {\bibinfo {volume} {97}},\ \bibinfo {pages}
  {221102} (\bibinfo {year} {2006}{\natexlab{a}})}\BibitemShut {NoStop}%
\bibitem [{\citenamefont {{Abramowski}}\ \emph {et~al.}(2011)\citenamefont
  {{Abramowski}} \emph {et~al.}}]{NekrassovEldik:2010}%
  \BibitemOpen
  \bibfield  {author} {\bibinfo {author} {\bibfnamefont {F.}~\bibnamefont
  {{Abramowski}}} \emph {et~al.} (\bibinfo {collaboration} {H.E.S.S.
  Collaboration}),\ }\href {\doibase 10.1111/j.1365-2966.2009.16014.x}
  {\bibfield  {journal} {\bibinfo  {journal} {Phys. Rev. Lett.}\ }\textbf
  {\bibinfo {volume} {106}},\ \bibinfo {pages} {161301} (\bibinfo {year}
  {2011})}\BibitemShut {NoStop}%
\bibitem [{\citenamefont {{Aleksic}}(2011)}]{MagicSegue:2011}%
  \BibitemOpen
  \bibfield  {author} {\bibinfo {author} {\bibfnamefont {J.}~\bibnamefont
  {{Aleksic}}},\ }in\ \href@noop {} {\emph {\bibinfo {booktitle} {International
  Cosmic Ray Conference}}},\ \bibinfo {series} {International Cosmic Ray
  Conference}, Vol.~\bibinfo {volume} {5}\ (\bibinfo {year} {2011})\ p.\
  \bibinfo {pages} {149}\BibitemShut {NoStop}%
\bibitem [{\citenamefont {{Aliu}}\ \emph {et~al.}(2012)\citenamefont {{Aliu}}
  \emph {et~al.}}]{VeritasSegue:2012}%
  \BibitemOpen
  \bibfield  {author} {\bibinfo {author} {\bibfnamefont {E.}~\bibnamefont
  {{Aliu}}} \emph {et~al.},\ }\href {\doibase 10.1103/PhysRevD.85.062001}
  {\bibfield  {journal} {\bibinfo  {journal} {\prd}\ }\textbf {\bibinfo
  {volume} {85}},\ \bibinfo {eid} {062001} (\bibinfo {year}
  {2012})}\BibitemShut {NoStop}%
\bibitem [{\citenamefont {Hill}\ \emph {et~al.}(1987)\citenamefont {Hill},
  \citenamefont {Schramm},\ and\ \citenamefont {Walker}}]{Hill:1987}%
  \BibitemOpen
  \bibfield  {author} {\bibinfo {author} {\bibfnamefont {C.~T.}\ \bibnamefont
  {Hill}}, \bibinfo {author} {\bibfnamefont {D.~N.}\ \bibnamefont {Schramm}}, \
  and\ \bibinfo {author} {\bibfnamefont {T.~P.}\ \bibnamefont {Walker}},\
  }\href {\doibase 10.1103/PhysRevD.36.1007} {\bibfield  {journal} {\bibinfo
  {journal} {Phys. Rev. D}\ }\textbf {\bibinfo {volume} {36}},\ \bibinfo
  {pages} {1007} (\bibinfo {year} {1987})}\BibitemShut {NoStop}%
\bibitem [{\citenamefont {{Tasitsiomi}}\ and\ \citenamefont
  {{Olinto}}(2002)}]{Tasitsiomi:2002}%
  \BibitemOpen
  \bibfield  {author} {\bibinfo {author} {\bibfnamefont {A.}~\bibnamefont
  {{Tasitsiomi}}}\ and\ \bibinfo {author} {\bibfnamefont {A.~V.}\ \bibnamefont
  {{Olinto}}},\ }\href {\doibase 10.1103/PhysRevD.66.083006} {\bibfield
  {journal} {\bibinfo  {journal} {\prd}\ }\textbf {\bibinfo {volume} {66}},\
  \bibinfo {pages} {083006} (\bibinfo {year} {2002})}\BibitemShut {NoStop}%
\bibitem [{\citenamefont {Bogovalov}\ and\ \citenamefont
  {Aharonian}(2000)}]{PulsarLines:2000}%
  \BibitemOpen
  \bibfield  {author} {\bibinfo {author} {\bibfnamefont {S.~V.}\ \bibnamefont
  {Bogovalov}}\ and\ \bibinfo {author} {\bibfnamefont {F.~A.}\ \bibnamefont
  {Aharonian}},\ }\href {\doibase 10.1046/j.1365-8711.2000.03250.x} {\bibfield
  {journal} {\bibinfo  {journal} {MNRAS}\ }\textbf {\bibinfo {volume} {313}},\
  \bibinfo {pages} {504} (\bibinfo {year} {2000})}\BibitemShut {NoStop}%
\bibitem [{\citenamefont {{Bringmann}}\ \emph {et~al.}(2008)\citenamefont
  {{Bringmann}}, \citenamefont {{Bergstr{\"o}m}},\ and\ \citenamefont
  {{Edsj{\"o}}}}]{Bergstroem:2007}%
  \BibitemOpen
  \bibfield  {author} {\bibinfo {author} {\bibfnamefont {T.}~\bibnamefont
  {{Bringmann}}}, \bibinfo {author} {\bibfnamefont {L.}~\bibnamefont
  {{Bergstr{\"o}m}}}, \ and\ \bibinfo {author} {\bibfnamefont {J.}~\bibnamefont
  {{Edsj{\"o}}}},\ }\href {\doibase 10.1088/1126-6708/2008/01/049} {\bibfield
  {journal} {\bibinfo  {journal} {J. High Energy Phys.}\ }\textbf {\bibinfo
  {volume} {1}},\ \bibinfo {pages} {49} (\bibinfo {year} {2008})}\BibitemShut
  {NoStop}%
\bibitem [{\citenamefont {Ackermann}\ \emph {et~al.}(2012)\citenamefont
  {Ackermann} \emph {et~al.}}]{FermiLines:2012}%
  \BibitemOpen
  \bibfield  {author} {\bibinfo {author} {\bibfnamefont {M.}~\bibnamefont
  {Ackermann}} \emph {et~al.} (\bibinfo {collaboration} {FERMI-LAT
  Collaboration}),\ }\href {\doibase 10.1103/PhysRevD.86.022002} {\bibfield
  {journal} {\bibinfo  {journal} {Phys. Rev. D}\ }\textbf {\bibinfo {volume}
  {86}},\ \bibinfo {pages} {022002} (\bibinfo {year} {2012})}\BibitemShut
  {NoStop}%
\bibitem [{\citenamefont {{Bringmann}}\ \emph {et~al.}(2012)\citenamefont
  {{Bringmann}}, \citenamefont {{Huang}}, \citenamefont {{Ibarra}},
  \citenamefont {{Vogl}},\ and\ \citenamefont {{Weniger}}}]{Bringmann:2012}%
  \BibitemOpen
  \bibfield  {author} {\bibinfo {author} {\bibfnamefont {T.}~\bibnamefont
  {{Bringmann}}}, \bibinfo {author} {\bibfnamefont {X.}~\bibnamefont
  {{Huang}}}, \bibinfo {author} {\bibfnamefont {A.}~\bibnamefont {{Ibarra}}},
  \bibinfo {author} {\bibfnamefont {S.}~\bibnamefont {{Vogl}}}, \ and\ \bibinfo
  {author} {\bibfnamefont {C.}~\bibnamefont {{Weniger}}},\ }\href {\doibase
  10.1088/1475-7516/2012/07/054} {\bibfield  {journal} {\bibinfo  {journal} {J.
  Cosm. Astrop. Phys.}\ }\textbf {\bibinfo {volume} {7}},\ \bibinfo {eid} {054}
  (\bibinfo {year} {2012})}\BibitemShut {NoStop}%
\bibitem [{\citenamefont {{Weniger}}(2012)}]{Weniger:2012}%
  \BibitemOpen
  \bibfield  {author} {\bibinfo {author} {\bibfnamefont {C.}~\bibnamefont
  {{Weniger}}},\ }\href {\doibase 10.1088/1475-7516/2012/08/007} {\bibfield
  {journal} {\bibinfo  {journal} {J. Cosm. Astrop. Phys.}\ }\textbf {\bibinfo
  {volume} {8}},\ \bibinfo {eid} {007} (\bibinfo {year} {2012})}\BibitemShut
  {NoStop}%
\bibitem [{\citenamefont {{Tempel}}\ \emph {et~al.}(2012)\citenamefont
  {{Tempel}}, \citenamefont {{Hektor}},\ and\ \citenamefont
  {{Raidal}}}]{Tempel:2012}%
  \BibitemOpen
  \bibfield  {author} {\bibinfo {author} {\bibfnamefont {E.}~\bibnamefont
  {{Tempel}}}, \bibinfo {author} {\bibfnamefont {A.}~\bibnamefont {{Hektor}}},
  \ and\ \bibinfo {author} {\bibfnamefont {M.}~\bibnamefont {{Raidal}}},\
  }\href {\doibase 10.1088/1475-7516/2012/09/032} {\bibfield  {journal}
  {\bibinfo  {journal} {J. Cosm. Astrop. Phys.}\ }\textbf {\bibinfo {volume}
  {9}},\ \bibinfo {eid} {032} (\bibinfo {year} {2012})}\BibitemShut {NoStop}%
\bibitem [{\citenamefont {{Bergstr{\"o}m}}\ \emph {et~al.}(2001)\citenamefont
  {{Bergstr{\"o}m}}, \citenamefont {{Edsj{\"o}}},\ and\ \citenamefont
  {{Ullio}}}]{Bergstrom:2001}%
  \BibitemOpen
  \bibfield  {author} {\bibinfo {author} {\bibfnamefont {L.}~\bibnamefont
  {{Bergstr{\"o}m}}}, \bibinfo {author} {\bibfnamefont {J.}~\bibnamefont
  {{Edsj{\"o}}}}, \ and\ \bibinfo {author} {\bibfnamefont {P.}~\bibnamefont
  {{Ullio}}},\ }\href {\doibase 10.1103/PhysRevLett.87.251301} {\bibfield
  {journal} {\bibinfo  {journal} {Phys. Rev. Lett.}\ }\textbf {\bibinfo
  {volume} {87}},\ \bibinfo {pages} {251301} (\bibinfo {year}
  {2001})}\BibitemShut {NoStop}%
\bibitem [{\citenamefont {Pieri}\ \emph {et~al.}(2011)\citenamefont {Pieri},
  \citenamefont {Lavalle}, \citenamefont {Bertone},\ and\ \citenamefont
  {Branchini}}]{Pieri:2011}%
  \BibitemOpen
  \bibfield  {author} {\bibinfo {author} {\bibfnamefont {L.}~\bibnamefont
  {Pieri}}, \bibinfo {author} {\bibfnamefont {J.}~\bibnamefont {Lavalle}},
  \bibinfo {author} {\bibfnamefont {G.}~\bibnamefont {Bertone}}, \ and\
  \bibinfo {author} {\bibfnamefont {E.}~\bibnamefont {Branchini}},\ }\href
  {\doibase 10.1103/PhysRevD.83.023518} {\bibfield  {journal} {\bibinfo
  {journal} {Phys. Rev. D}\ }\textbf {\bibinfo {volume} {83}},\ \bibinfo
  {pages} {023518} (\bibinfo {year} {2011})}\BibitemShut {NoStop}%
\bibitem [{\citenamefont {Aharonian}\ \emph
  {et~al.}(2006{\natexlab{b}})\citenamefont {Aharonian} \emph
  {et~al.}}]{Crab:2006}%
  \BibitemOpen
  \bibfield  {author} {\bibinfo {author} {\bibfnamefont {F.}~\bibnamefont
  {Aharonian}} \emph {et~al.} (\bibinfo {collaboration} {H.E.S.S.
  Collaboration}),\ }\href@noop {} {\bibfield  {journal} {\bibinfo  {journal}
  {Astron. Astrophys.}\ }\textbf {\bibinfo {volume} {457}},\ \bibinfo {pages}
  {899} (\bibinfo {year} {2006}{\natexlab{b}})}\BibitemShut {NoStop}%
\bibitem [{\citenamefont {James}\ and\ \citenamefont
  {Roos}(1975)}]{James:1975dr}%
  \BibitemOpen
  \bibfield  {author} {\bibinfo {author} {\bibfnamefont {F.}~\bibnamefont
  {James}}\ and\ \bibinfo {author} {\bibfnamefont {M.}~\bibnamefont {Roos}},\
  }\href {\doibase 10.1016/0010-4655(75)90039-9} {\bibfield  {journal}
  {\bibinfo  {journal} {Comput.Phys.Commun.}\ }\textbf {\bibinfo {volume}
  {10}},\ \bibinfo {pages} {343} (\bibinfo {year} {1975})}\BibitemShut
  {NoStop}%
\bibitem [{\citenamefont {James}(2004)}]{James:2004}%
  \BibitemOpen
  \bibfield  {author} {\bibinfo {author} {\bibfnamefont {F.}~\bibnamefont
  {James}},\ }\href@noop {} {\bibfield  {journal} {\bibinfo  {journal}
  {http://seal.cern.ch/documents/minuit/mnerror.ps}\ } (\bibinfo {year}
  {2004})}\BibitemShut {NoStop}%
\bibitem [{\citenamefont {{de Naurois}}\ and\ \citenamefont
  {{Rolland}}(2009)}]{Naurois:2009}%
  \BibitemOpen
  \bibfield  {author} {\bibinfo {author} {\bibfnamefont {M.}~\bibnamefont {{de
  Naurois}}}\ and\ \bibinfo {author} {\bibfnamefont {L.}~\bibnamefont
  {{Rolland}}},\ }\href {\doibase 10.1016/j.astropartphys.2009.09.001}
  {\bibfield  {journal} {\bibinfo  {journal} {Astrop. Phys.}\ }\textbf
  {\bibinfo {volume} {32}},\ \bibinfo {pages} {231} (\bibinfo {year}
  {2009})}\BibitemShut {NoStop}%
\end{thebibliography}%

\end{document}